\documentstyle[]{mn}

\newlength{\colwidthf}
\newlength{\hcolwidthf}
\newlength{\colwidth}
\newlength{\hcolwidth}
\input ./psfig.sty
\newif\ifpsfiles\psfilestrue
\psfilestrue
\def\getfig#1#2{\ifpsfiles\psfig{figure=#1,width=\hsize}\else\vskip#2\fi}

\def\kpc{\,{\rm kpc}}

\def\spose#1{\hbox to 0pt{#1\hss}}
\def\lta{\mathrel{\spose{\lower 3pt\hbox{$\mathchar"218$}}
     \raise 2.0pt\hbox{$\mathchar"13C$}}}
\def\gta{\mathrel{\spose{\lower 3pt\hbox{$\mathchar"218$}}
     \raise 2.0pt\hbox{$\mathchar"13E$}}}
\def\pc{{\rm\,pc}}
\def\kpc{{\rm\,kpc}}

\title[Parametric models in noisy inhomogeneous regression problems]
{A graphical selection method for parametric models in noisy 
inhomogeneous
regression}

\author[N. Bissantz and A. Munk]{Nicolai Bissantz and Axel 
Munk\\
Institut f\"ur Mathematische Stochastik, Universit\"at G\"ottingen, Lotzestr. 13,
37083 G\"ottingen, Germany}
\def\3{\ss}

\newcommand{\bea}{\begin{eqnarray*}}
\newcommand{\eea}{\end{eqnarray*}}
\newcommand{\be}{\begin{eqnarray}}
\newcommand{\ee}{\end{eqnarray}}
\begin{document}
\maketitle

\begin{abstract}
A common problem in physics is to fit regression data by a parametric class
of functions, and to decide whether a certain functional form allows for
a good fit of the data. Common goodness of fit methods are based on 
the calculation of the distribution of certain statistical quantities 
under the assumption that the model under consideration {\it holds true}. 
This proceeding bears methodological flaws, e.g. a good ``fit'' - albeit
the model is wrong - might be due to over-fitting, or to the fact
that the chosen statistical criterion is not powerful 
enough against the present particular deviation between 
model and true regression function.
This causes particular difficulties when models with different numbers of
parameters are to be compared. Therefore the number of parameters
is often penalised additionally. We provide a 
methodology which circumvents these problems to some extent. It is based on
the consideration of the error distribution of the goodness of fit
criterion under a broad range of possible models - and not only under
the assumption that a given model holds true.
We present a graphical method to decide for the most evident model
from a range of parametric models of the data. The method allows to 
quantify statistical evidence {\it for} the model (up to some distance
between model and true regression function) and not only {\it absence 
of evidence} against, as common goodness of fit methods
do. Finally we apply our method 
to the problem of recovering
the luminosity density of the Milky Way from a de-reddened {\it COBE/DIRBE}
L-band map. We present statistical evidence for flaring of the 
stellar disc inside the solar circle.
\end{abstract}


\section{Introduction}
Often one is confronted with the problem to reconstruct an 
unknown function $f(t_i)$ from noisy observations 
$y_i=y(t_i), i=1, \ldots, N$.
Astrophysical examples include reverberation mapping 
of gas in active galactic nuclei and recovery of  the spatial 
(three-dimensional) luminosity density of a galaxy from 
blurred observations of its surface brightness. See 
e.g. Lucy (1994) for more examples of astronomical inverse
problems. 
In this paper we are concerned with a new method to compare
several competing parametric models for the regression function $f$. 

Due to the noisy measurements it is tempting to assume that
$y_{i}=f(t_{i})+\varepsilon_{i},$
where the $\varepsilon_{i}$ denote some random
noise and $f(t_{i})$ the expected value of $y_{i}$, i.e. 
$E[y_{i}]=f(t_i).$
In particular we allow for different error distributions of the 
$\varepsilon_{i}$, which entails inhomogeneous variance patterns,
viz. $V[\varepsilon_{i}]=\sigma_{i}^2$, as will be the case in our example
of de-projecting the de-reddened {\it COBE/DIRBE} L-band surface brightness 
map of Spergel et al. (1996), as discussed by Bissantz \& Munk (2001, 
[BM1]). 

It is a common proceeding to fit a class of functions $U=\{f_{\vartheta}:
\vartheta\in\Theta\}$ (parametric model)
to the data $y_i$. The parametric model may depend on a parameter $\vartheta$,
where $\vartheta\in\Theta \subseteq I\!\!R^d$. A popular method to select a 
``best-fitting'' $\vartheta$
from $\Theta$ is to minimise the empirical mean squared error (MSE)
\be
Q^2_N(\vartheta):=\sum_{i=1}^{N} \left(y_i-f_{\vartheta}(t_i)\right)^2
\label{eq1}
\ee
or weighted variants of it. This gives $\hat{\vartheta}$, the least 
squares estimator (LSE) of $\hat{\vartheta}$. Other measures of 
goodness of fit are e.g. $L^1$-error criteria, where the absolute 
deviation between $y_i$ and $f_{\vartheta}(t_i)$ is considered (Seber \&
Wild 1989). A small value of $Q_N^2(\hat{\vartheta})$ often is used as
an indication for a good explanation of the observations by the model 
$f_{\hat{\vartheta}}$. 
Note that in regression models where the noise is {\it inhomogeneous}
the quantity $Q_N^2(\hat{\vartheta})$ is often 
useless (cf. [BM1] for an explanation) and more subtle 
methods have to be applied. 

An advantage of the parametric fitting methodology
in contrast to nonparametric curve 
estimation, i.e. approximating the data by arbitrary functions (e.g. 
by splines, orthogonal series or wavelets, cf. Efromovich, 1999 or 
Hart, 1997) is the
fact that often physical reasoning resulting from a theory suggests
such a class of functions $U$. 
Furthermore, subsequent data analysis and interpretation becomes
very simple if once a proper $f_{\hat{\vartheta}}$ is selected.
Hence it is an important task to pre-specify 
$U$ correctly in order to obtain a reasonable fit.
Therefore in this paper we discuss the problem of evaluating the goodness
of fit of a parametric model $U$. Moreover, we offer a graphical method
which allows to select a proper model $U$ from a class of different 
models ${\cal U}=\{U_k\}_{k=1,\ldots,l}$, say. 

A common proceeding is to assume that the model holds,
and to test if the observed data give reason to reject the model. 
This type of goodness of fit tests is performed 
by evaluating the probability distribution
of a pre-specified measure of discrepancy, such as $Q_N^2(\hat\vartheta)$. 
This is done under the assumption that $U$ holds true. 
Then, when this measure exceeds a certain quantity,
the model $U$ is rejected. 

One problem of such methods is that a
large data set leads essentially to rejection of any
model $U$ (an illustrative discussion can be found in Berger, 1985),
because the ``real world'' is never exactly
described by such a model and as the number of observations increases,
statistical methods will always detect these deviations between the model
and ``reality''. 
Conversely, the selected statistical criterion may lead to
a decision in favour of $U$ (albeit wrong), because it is not capable 
to detect important deviations from $U$ or the decision is affected by 
quantities which are
not captured in the model $U$ (e.g. correlation between the $y_{i}$).
Another problem can be over-fitting of data by models with a too large number
of parameters. Therefore, various methods have been suggested which penalise the
number of parameters, i.e. the complexity of a model (Akaike, 1974,
Burnham et al., 1998, or Schwarz, 1978). 

In this paper, we suggest a 
methodology which aims to avoid these problems by considering 
the distribution of a discrepancy measure such as $Q_N^2(\hat{\vartheta})$
under all ``possible'' functions $f$. This extends the method given in 
[BM1] to the more realistic case where the ``true'' function 
$f$ is not restricted to be in $U$.
Furthermore, a graphical method will be presented which allows to select the most 
appropriate between several competing models $U_i$. With our method, 
this is still possible if these models have different numbers of parameters. 

In the next section we will describe the method and its algorithmic
implementation, the wild bootstrap.  Based on the theory presented 
in sect. 2, we suggest in sect. 3 a graphical method to assess
the validity of $U$ as well as to compare between different models.
This method is denoted as $p$-value curve analysis. In sect. 4 our 
method is applied to a near-infrared [NIR] L-band map of the Milky Way [MW]
and two different models of the spatial luminosity distribution are compared. 
One of the models includes a flaring disc component. 
We analyse the models' $p$-value curves, and find that  
flaring in the disc improves the fit to the data. 

\section{A new method of model selection}
In section 2.1  we briefly recall the methodology suggested in [BM1]
and extend it to the situation where $f$ is not in the model $U$.
This will be used to compute $p$-value curves, a graphical method
of model diagnostics, which was introduced by Munk \& Czado (1998) in
a different context. 
In sect. 2.2 we describe the practical application of the method.

\subsection{Basic theory of the method}
We begin with an introduction to the basic principles of our method.
As mentioned above $Q^2_N(\hat\vartheta)$ fails to be a valid criterion 
for goodness of fit in inhomogeneous models [BM1]. Instead we replace the 
pure residuals
$y_{i}-f_{\hat\vartheta}(t_{i})$ with smoothed residuals, to allow
for a valid statistical analysis. For the smoothing step we require 
an {\it injective} linear integral operator with kernel $T$, viz.
\bea
g(w)={\bf T}(f)(w)=\int T(w,v)f(v)dv
\eea
which maps the function $f$ to be recovered onto $g$. In principle any
injective operator $\bf T$ is a valid option for the smoothing, however
a good choice is driven by aspects such as efficiency and simplicity.
In our example (cf. sect. 4) we introduce  
``cumulative smoothing'' 
with $T(w,v)={\rm min}(w,v)$. An extensive simulation
study by Munk \& Ruymgaart (1999) revealed this smoothing kernel as
a reasonable choice which yields a procedure capable to detect a broad 
range of deviations from the class of functions $U=\{f_{\vartheta}:
\vartheta\in\Theta\}$. 

A measure of the discrepancy between the ``true'' $f$ and $U$ is the 
transformed distance 
\be
D^2(f)=\min_{\vartheta\in\Theta} ||{\bf T}(f-f_{\vartheta})||^2
\label{eq2}
\ee
where the norm refers to some $L^2$-norm.
Now assume that the minimum in eq. \ref{eq2} is achieved at a parameter vector
$\vartheta^{\ast}=\vartheta^{\ast}(g)\in\Theta$. Because $\vartheta^{\ast}$ is
unknown it has to be estimated from the data. This can be done by numerical
minimisation of the empirical counterpart of the r.h.s. of eq. \ref{eq2},
\bea
\hat{D}^2:=\min_{\vartheta\in\Theta} ||{\bf T}f_{\vartheta}-\hat{g}||^2 
\eea
where
\bea
\hat{g}=N^{-1} \sum_{i=1}^{N} y_{i} T(u,t_{i})
\eea
is an estimation of $g$ using the noisy data $y_i$. The reasoning behind this
approach is that for sufficiently large number $N$ of observations (in our
example $N=4800$) it can be shown that $\hat g$ converges in probability to the
true (but unknown) function $g$, independently whether a parametric model
$U$ is valid or not. On the other hand the empirical minimiser $\hat\vartheta_T$
estimates the best possible fit to $\hat g$ by the model $U$. 
The resulting estimator is denoted as a smoothed minimum distance estimator
$\hat{\vartheta}_{\bf T}$ (SMDE) and has the property that, if the true function
$f=g_{\vartheta^{\ast}}$ is in $U$, $\hat{\vartheta}_{\bf T}\rightarrow 
\vartheta^{\ast}$ as the sample size increases. For detailed proofs we refer to
Munk \& Ruymgaart (1999). 

Note that $\vartheta^{\ast}$ is the ``true'' best-fitting
parameter vector, which could only be determined if the data would be free of noise,
whereas $\hat{\vartheta}_{\bf T}$ is an estimation of the best-fitting parameter vector
using the noisy data. Here and in the following, quantities with a hat, 
`` $\hat{ }$ '', are estimated from
the noisy data, whereas such without a hat are the ``true'' functions to be recovered. 

Munk \& Ruymgaart (1999) showed that the probabilistic
limiting behaviour of $\hat{D}^2$ depends on whether $f$ belongs to the 
model $U$ under investigation. More precisely when $f$ belongs to $U$ the 
distribution of $N\hat D^2$ is for large $N$ approximately that of  
\be
\sum_{i=1}^{\infty} \lambda_i \chi_i^2 
\label {eqa}
\ee
where $\chi_i^2$ denotes a sequence of independent squares of standard normal
random variables and $\lambda_i\geq 0$ is a sequence of real numbers, s.t.
$\sum_{i=1}^{\infty}\lambda^2_i<\infty$, which depend on $\vartheta^{\ast}$, 
the distribution
${\cal L}$ of errors $\varepsilon_i$ and the operator $T$. 

In contrast if $f$ does not belong to $U$, we have $D^2>0$ and
$N^{\frac{1}{2}}\left(\hat{D}^2-D^2\right)$
tends for large $N$ to a centred normal distribution with variance 
$\sigma^2_{{\bf T},{\cal L},\vartheta^{\ast}}$, depending on $\bf T$, $\vartheta^{\ast}$,
and ${\cal L}$. Observe,
that we obtain two different types of distributions, accordingly to the
situation whether the ``true'' (unknown) function $f$ is in the model $U$ or
not. Because of the complicated dependency of the $(\lambda_i)_{i\in I\!\!N}$
and $\sigma^2_{{\bf T},{\cal L},\vartheta^{\ast}}$ on $\bf T$, $\vartheta^{\ast}$,
and ${\cal L}$ a resampling algorithm
should be applied in order to approximate these limiting distributions.
Stute et al. (1998) presented a wild bootstrap algorithm which can be used
to approximate the law $N\hat D^2$.  Munk (1999) showed that this 
algorithm is also valid when $f$ does not belong to $U$, which is crucial for our 
paper. This algorithm will be carefully explained in the next paragraph. Recall
that the subsequent bootstrap algorithm allows to determine the probability 
distribution of the quantity of interest $\hat D^2$. The general strategy of our
method will be the following. Because $\hat D^2$ measures the distance between the
model $U$ and the estimator $\hat g$ from noisy data, knowledge of the probability
distribution of $\hat D^2$ (which will be determined by the subsequent bootstrap
algorithm) allows us to quantify whether an observed value of $\hat D^2$ for a 
model $U$ is more likely than for a competing model $U'$, say. Even, when none of 
these models is completely true (which is always the case in the real world)
$\hat D^2$ quantifies the best possible {\sl approximation} of $g$ by $U$ or
$U'$ respectively.

\subsection{Practical application of the method}
We now introduce the resampling algorithm to approximate the law $N\hat D^2$.
The algorithm starts with the determination of the SMDE 
$\hat\vartheta_{\bf T}$ and the smoothed residuals between this model and the 
data (step 1). Then in step 2-5 the resampling part of the algorithm follows. 
The same algorithm is used in [BM1].

{\bf Step 1:}  ({\it Generate residuals}). Compute residuals
\bea
\hat \varepsilon _i
:= y_i-f_{\hat \vartheta_{\bf T}}
(t_i), \qquad i=1, \cdots, n
\eea
where ${\hat \vartheta_{\bf T}}$
denotes a solution of the
minimisation of
\bea
 \hat D^2 := \chi^2(\hat \vartheta_{\bf T})
:= \min_{\vartheta \in \Theta}
\| \hat g -  {\bf T}f_{\vartheta} \|^2.
\eea

\begin{figure}
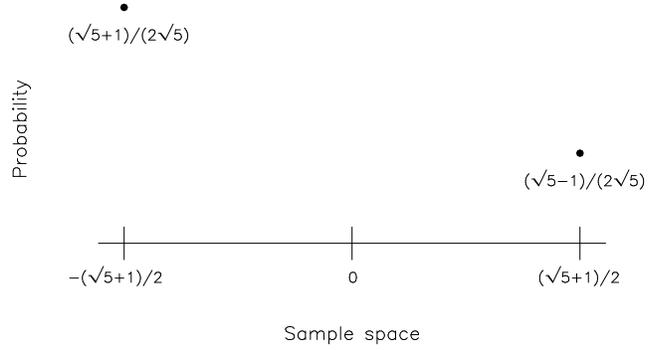

\getfig{figure0.eps}{6cm}
  \caption{Binary probability distribution required in step 2
of the wild bootstrap algorithm. The ordinate gives the 
probability of the random number to be $-(\sqrt{5}+1)/2$ 
and $(\sqrt{5}+1)/2$, respectively.}
\label{figprob}
\end{figure}

{\bf Step 2:} ({\it The "wild" part}).
Generate  new random variables $c_i^{\ast}, \; i=1, \ldots, n$, 
which do {\it not} depend on the data, where each $c_i^{\ast}$
is distributed to a distribution
which assigns probability
$(\sqrt{5}+1)/2 \sqrt{5} $ to the
value $(- \sqrt{5} -1)/2$ and
$(\sqrt{5} - 1)/2 \sqrt{5}$ to the
value $(\sqrt{5} + 1)/2$. See fig. \ref{figprob} for
a visualisation of this probability distribution.

{\bf Step 3:} ({\it Bootstrapping residuals}). Compute
$\varepsilon_i^{\ast} := \hat
\varepsilon_i c_i^{\ast}$ and $y_i^{\ast} = 
f_{\hat \vartheta_{\bf T}}
 + \varepsilon_i^{\ast}$.
This gives a new data vector 
$(y_i^{\ast}, t_i)_{i=1, \ldots, n}$.

{\bf Step 4:} ({\it Compute the target}).
Compute $\hat D^{2}_{\ast}$
with $(y_i^{\ast}, t_i)_{i=1, \ldots, n}$

{\bf Step 5:} ({\it Bootstrap replication}). Repeat step 1-4 {\it B}
times which gives values $\hat D_{1,\ast}^{2}, \ldots,
\hat D_{B,\ast}^{2}$. $B$ is a large number, typically $B=500$ or 
$B=1000$ is sufficient. \\

From the bootstrap replications $\hat D_{1,\ast}^{2}, \ldots,
\hat D_{B,\ast}^{2}$ we compute the quantities 
\bea
x_1=\sqrt{N}\left(\hat D_{1,\ast}^{2}-\hat D^2\right),\ldots, 
x_B=\sqrt{N}\left(\hat D_{B,\ast}^{2}-\hat D^2\right),
\eea
using the number
of data points $N$. The $x_1,\ldots,x_B$ are realisations
of the random quantity $X=\sqrt{N} \left( \hat D_{\ast}^{2}-\hat D^2\right)$.
It can be proved that the empirical distribution function of 
$\hat D^2_{1,\ast},\ldots,\hat D^2_{B,\ast}$ yields an approximation 
to the true distribution
of $\hat D^2$ after a proper re-centring, i.e. the cumulative probability
distribution function $F_{B}^{\ast}$ of 
$X=\sqrt[]{N}\left(\hat D^2_{\ast}-\hat D^2\right)$ is close to the cumulative
distribution function
of $\sqrt[]{N}\left(\hat D^2-D^2\right)$ for any $D^2> 0$ (Munk, 1999). 

An important application of this result is 
to determine an approximation to the probability $p(t, D^2)$ 
that $\hat D^2$ is below a certain value $t$, provided the distance between 
true function $f$ and the model is $D^2$.
To this end we use that $F_{B}^{\ast}$, 
found from the bootstrap replications,
approximates the (unknown) cumulative probability distribution of 
$\sqrt{N}(\hat D^2-D^2)$. 
The latter distribution allows to determine $p(t, D^2)$. 
Hence we are in the position 
to compare the probability that the observed value of $\hat D^2$ is 
achieved in all "possible worlds", i.e. for any possible $f$. In fact,
it turns out that this probability does only depend on $f$ via $D^2(f)$,
which allows a nice geometric interpretation as we will illustrate in
the following. 

We will use the asymptotic similarity of the
two cumulative probability distributions in the following section  
to estimate the probability $p(t,D^2)$. From this we then
define the $P$-value curve $\alpha_N(\Pi)$, which can be regarded 
as a measure of evidence for $D^2\leq \Pi$, given $\hat D^2$ and $F_{B}^{\ast}$.
Thus these quantities allow to constrain $D^2$ for a
parametric model of a given set of data.

\begin{figure*}
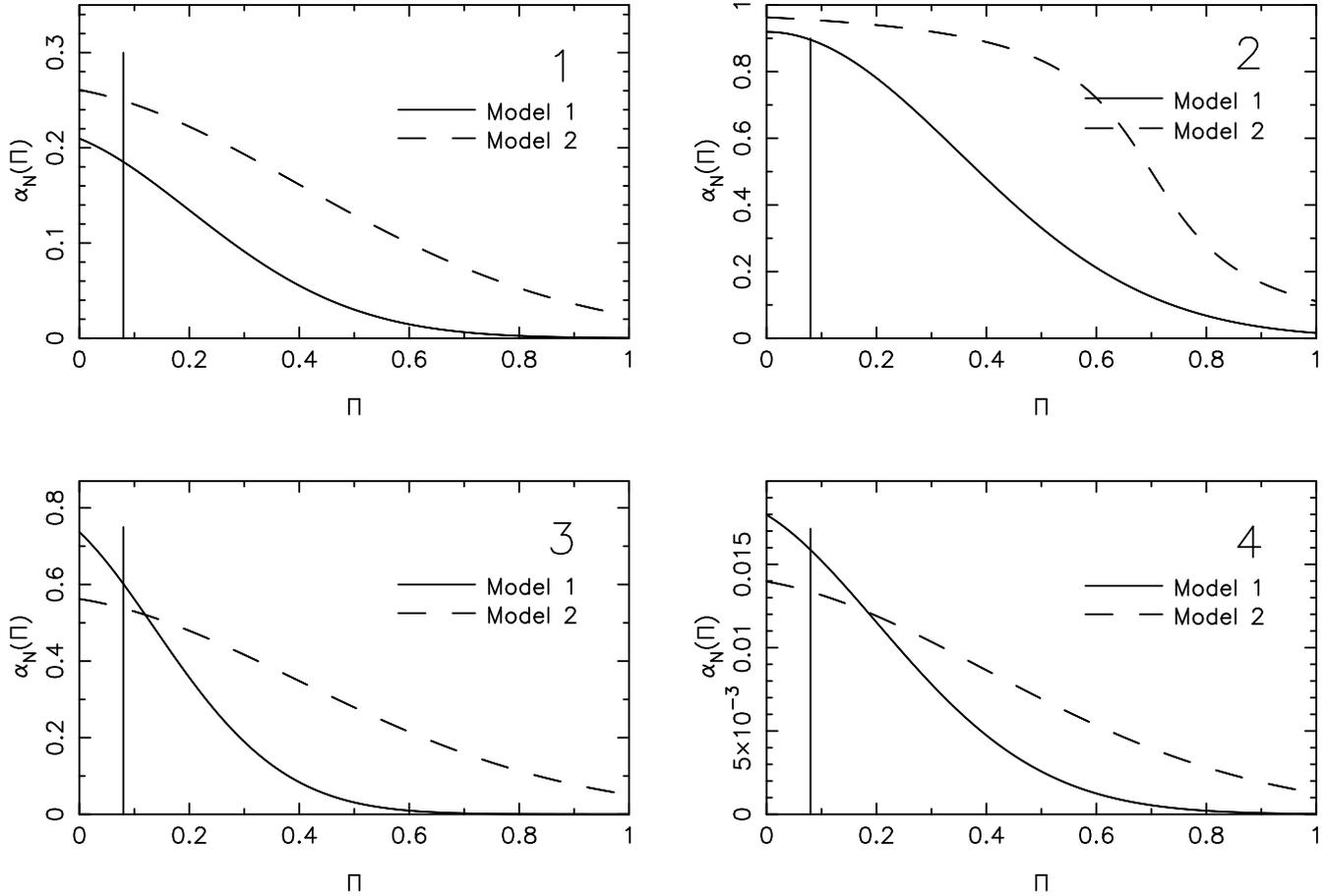

\getfig{prinzipfig.eps}{6cm}
  \caption{Typical cases for $p$-value curve comparison of two parametric
models. The vertical lines at $\Pi=0.08$ in the graphs indicate the 
observed value of
$\hat D^2=0.08$. In graph $1$, model $1$ fits better, as well as in graph $2$. 
However
in graph $2$ is additionally strong evidence that model $1$ does not hold. 
Graph  $3$ is again an example with model $1$ the better model.
Finally in graph $4$ the situation depends on the assumption of the distance
between the parametric model $U$ and the true regression function $f$ (cf. sect. 2) 
$D^2$.}
\label{figbeispielfig}
\end{figure*}

\section{$P$-value curves}
The main methodology we propose in this paper is the computation
of a $p$-value curve as a graphical tool for illustrating the evidence of a model. 
To this end we plot the function $\alpha_N(\Pi)=F^{\ast}_B\left(\sqrt[]{N}
\left(\hat{D}^2-\Pi\right)\right)$ for $\Pi>0$, i.e. the value of $\alpha_N(\Pi)$
is given by the probability that the random quantity 
$X=\sqrt{N}\left(\hat D_{\ast}^2-
\hat D^2\right)$ is smaller than $\sqrt{N}\left(\hat D^2-\Pi\right)$. 
Note that this implies that for $\Pi$ increasing $\alpha_N(\Pi)$ decreases,
because we then evaluate the cumulative distribution function $F_B^{\ast}(x)$
for decreasing $x$, and in particular, if $\alpha_N(\Pi)$ is small, 
at the left tail of $F_B^{\ast}$. 

The interpretation of the function $\alpha_N(\Pi)$ is as follows.
Assume the true distance between model $U$ and $f$ (i.e. 
the distance between the minimising $f_{\vartheta^{\ast}}$ and the 
``true'' function $f$) is $D^2=\Pi$. If this holds, 
the probability that $\sqrt[]{N}(\hat{D}^2-D^2)$ is smaller than some value $t$
is given as
\be
P_{D^2=\Pi}\left(\sqrt[]{N}\left(\hat{D}^2-D^2\right)\leq t\right)\approx F^{\ast}_B(t)
\ee
where the r.h.s. denotes the bootstrap approximation to the true distribution
function on the l.h.s. 
Now we reject the hypotheses $H:D^2>\Pi$ (vs. alternative $K:D^2\leq \Pi$)
whenever $\alpha_N(\Pi)\leq \alpha$ for a given level of significance $\alpha$.
Hence $1-\alpha_N(\Pi)$ can be regarded as the estimated evidence in favour of the
model $U$ (up to a distance between model and data $D^2\leq\Pi$). 

Note that this approach highlights the fact that finally the astrophysicist
has to decide whether a value of $D^2=\Pi$ should be regarded as scientifically
negligible
or as deviation from the model $U$ which is considered as too large by astrophysical
reasons. We mention that the classical goodness of fit tests
do not offer the scientist the specification of such a value $\Pi$. 

How can an upper bound for a just acceptable $D^2$ be determined? 
One simple suggestion is to 
compute the distance $\tilde D^2=||{\bf T}f_{\hat\vartheta_{\bf T}}-{\bf T}\tilde 
f_{\hat\vartheta_{\bf T}}||^2$ between the best model $f_{\hat\vartheta_{\bf T}}$ 
and "test models" $\tilde f_{\hat\vartheta_{\bf T}}$. 
Such test models should then be constructed from $f_{\hat\vartheta_{\bf T}}$ 
by adding (systematic) deviations to the model, which are
still considered as scientifically negligible differences to the best model.
Then, if $D^2$ is not larger than the average over the test models 
$<\!\tilde D^2\!>$, computed from a number of such test
models, it is considered as scientifically negligible.

Observe that with our proposed method the statistical type one error is the error
to decide for the model (or more precise for a neighbourhood $D^2\leq \Pi$ of the 
model) although it is not valid. Classical goodness of fit tests are only able
to control the error of rejecting the model albeit it holds, i.e. they are based on 
testing $H_0: D^2=0$ vs. $K_0: D^2>0$.
 
Fixing $\hat{D}^2$, a small value of $\alpha_N(\Pi)$ indicates 
large probability for $D^2\leq\Pi$ and a large value (close to $1$) of 
$\alpha_N(\Pi)$ indicates a large probability for $D^2>\Pi$. It is important
to note that the interesting regions of the resulting curves $\alpha_N(\Pi)$ 
are those values of $\Pi$
where $\alpha_N(\Pi)$ is rather large (larger than $0.9$ say) and rather small
(smaller than $0.1$) in accordance with the usual choice of levels of significance.
In contrast decisions based on $\alpha_N$ in regions
where $\alpha_N(\Pi)\approx 0.5$ would correspond to flipping a coin in order to 
decide whether $D^2\leq\Pi$ or not. 

As an important advantage of $p$-value curves we find that it gives us not only
an estimated probability ($p$-value) that we would observe a test statistic (such as 
$Q_N^2(\hat{\vartheta})$ or $\hat{D}^2$) provided the assumption that $U$
underlies the data is true. Rather we obtain simultaneously all
scenarios over the entire range of ``possible worlds'' which are parametrised 
by $D^2$. In particular this implies that models with a large number of
parameters are penalised in an automatic way. As the number of parameters
increases the variability of the statistic $\hat{D}^2$ increases and hence
the variability of $F^{\ast}_B$, i.e. the range of values for $X$, for
which $F^{\ast}_B$ differs significantly from $0$ and $1$, is larger. On
the other hand the bias is reduced. As the number of parameters decrease 
the opposite will be the case.  This leads to a curve $\alpha_n(\Pi)$ which
slowly decreases to zero if the variance is too large or if the bias is too
large. Hence evidence for a small $\Pi$ can only be
claimed if these two quantities are balanced.  

In other words a $p$-value curve reflects automatically 
the tradeoff between variance and bias in a regression. 
Here the bias of the regression functions can be viewed 
as the difference between the ``true'' expectation value 
$E[y_i]$ and the value of the regression function $f(t_i)$. The 
variance provides an estimate of the uncertainty of the best-fitting
parameters $\hat\vartheta$ or $\hat\vartheta_{\bf T}$. 

Before we analyse two competing models for the structure of the
MW we illustrate in an artifical example typical features of $p$-value curves.
In fig. \ref{figbeispielfig} various scenarios are displayed.
In graph $1$ model $1$ beats model $2$ at all fronts. The estimated evidence
for $D^2\leq\Pi$ 
is uniformly larger for any $\Pi>0$. 
This coincides with ``classical testing''
because also the classical $p$-value
for testing H:$D^2=0$ is larger. 
Observe that the classical $p$-value corresponds
in this graph to $1-\alpha_N(0)$. 

Graph $2$ is similar, observe however, that a classical analysis would indicate
that here is additionally strong evidence that model $1$ does not hold 
($\alpha_N(0)\gta 0.9$),
although it yields a better fit as model $2$, exactly as in graph $1$.   
Here the value of $\Pi$ where $\alpha_N(\Pi)=0.1$ (i.e. where $\Pi\approx 0.7$), 
say, becomes important because it gives
an idea of the order of magnitude between model $U$ and the true regression.
Hence it has to be decided for the particular problem whether a distance of $\Pi\approx
0.7$ is considered as ``large'' or scientifically irrelevant.

Graph $3$ represents a typical case of over-fitting by model $2$. 
Classical reasoning would prefer model $2$ because $\alpha_N(0)$ is smaller 
and hence the classical $p$-value larger. 
However, we see that this is due to a lack of
power of the used test statistic, because the slope of the 
curve is very flat due to a large variability of the test statistic. Hence there is not 
much support for the decision $D^2\leq 0.5$, say, ($\alpha_N(0.5)\approx 0.3$) whereas
model $1$ yields $\alpha_N(0.5)\approx 0.03$. Thus there is strong evidence that 
the distance between model $1$ and the true regression curve is smaller than $0.5$,
say. 

Finally in graph $4$ both models are
acceptable with slight preference to model $2$ provided a distance of $0.2$
(the point of intersection of both curves) is considered as an acceptable 
distance between $U$ and $f$. If a larger distance, $\Pi=0.5$, say is considered
to be tolerable, however model $1$ has to be preferred.

\section{Flaring of the stellar disc}
Observations have shown that the HI disc of the MW flares (see, for example,
Merrifield, 1992, or Malhotra, 1995). The situation is much less clear for the
stellar disc. Alard (2000) finds flaring for the disc outwards of the solar orbit, 
from an analysis of $2$ micron sky survey (2MASS) data, with 
a vertical scale-height $\approx 300\pc$ in the solar neighbourhood. 
Other evidence comes from Kent et al. (1991), who have fitted parametric
models to Spacelab2 IR telescope (IRT) $2.4\mu m$ observations of the MW.
The vertical scale-height of their best model's disc is constant in the inner 
$\approx 5\kpc$ with $h_z=165\pc$, but rises outside of this galactocentric radius 
to $h_z\approx 247\pc$ in the solar neighbourhood. Thus the results of Alard
and Kent et al. for the vertical disc scale-height in the solar neighbourhood 
are consistent to within $\approx 20\%$. 

We apply our proposed method to dust-corrected {\it COBE/DIRBE} L-band data 
(Weiland, 1994; Spergel et al., 1996), and  
investigate whether there is evidence for flaring of the disc inside the solar
orbit. We remark that this L-band observations are expected to trace the density 
of stars (Binney, Gerhard \& Spergel, 1997 [BGS]). Note that 
from non-parametric models of this L-band data [BGS] have found
vertical scale-heights $z_0\approx 120-150\pc$ at $R=5\kpc$ 
from the galactic centre. They remark that this
is inconsistent with the value of $300\pc$ from star counts at the Galactic poles
(Gilmore \& Reid, 1983), but consistent with the findings of Kent et al. (1991).
We will apply our proposed statistical test on the same data to
demonstrate its ability in this context as an example application.

The general outline of this sect. is as follows: First we introduce the
observational data (sect. 4.1), and construct functional forms for two different 
parametric models of the MW luminosity density distribution (sect. 4.2). 
Then (sect. 4.3)  we fit these models to the {\it COBE/DIRBE} L-band data and apply 
the wild bootstrap algorithm to both models, with $B= 5000$.
Finally we analyse the distribution of the distances $\hat D^2$ between the models and
the data, both under the assumption that the respective parametric model does reproduce
the data, and that this is not the case (sect. 4.4).

\subsection{Observational data}
The {\it DIRBE} experiment on board the {\it COBE} satellite, 
launched in {\it 1989},
has provided maps of the sky in several infrared wavebands
(Weiland et al. \cite{weiland94}).
This data has been used to estimate the luminosity distribution of the MW,
both parametrically (e.g. Freudenreich, 1998, and Dwek et al., 1995),
and non-parametrically (Binney \& Gerhard, 1996, [BGS], Bissantz et al., 1997,
and Bissantz \& Gerhard, 2001).

In this paper we use a {\it COBE/DIRBE} NIR L-band map, corrected for
dust absorption by Spergel et al. (\cite{spergel95}).  
The resolution of the equidistant grid of data is $n\! =\! 120$  points
in $-89.25\deg\! \leq\! l\! \leq \! 89.25\deg$ and $m\! =\! 40$ points in 
$-29.25\deg \! \leq\! b \!\leq\! 29.25\deg$. We only use the data 
$-60\deg\!\leq\! l \!\leq\! 60\deg$, $-20\deg\!\leq\! b \!\leq\! 10\deg$,
to downweight those parts of the sky where non-informative parts in the data 
can be observed due to extreme noise (cf. [BM1]). 

This dataset is well suited to demonstrate our proposed method since it consists
of several thousand data points, enough to make the method applicable.
Simulations have shown that the method is already applicable when more than $50$
data points are available provided the error distribution behaves well.

\subsection{The parametric models}
We construct two different parametric models, one including flaring, 
according to the approach of Kent et al. (1991), the other not.
In this section the functional forms of the models are presented, first
the individual bulge and disc components. We use a Cartesian coordinate system
with axes $x,y,z$. Here $x$ is along the major axis, and $y$ along the minor axis 
of the bulge/bar, both in the main plane of the MW. We set the position of the sun
in this coordinate system to a distance from the main plane of the disc 
$z_{\odot}=14\pc$, the distance to the galactic centre $R_{\odot}\!=\! 8\kpc$, 
and the angle between the major axis of the bar and the line-of-sight from the sun 
to the galactic centre $\phi_{\rm bar}=20\deg$ ([BGS]). Let
$a^2  \equiv  x^2+\left(\frac{y}{\eta}\right)^2 + 
\left(\frac{z}{\zeta}\right)^2$ and $r^2\equiv x^2+y^2$. Then the model 
components are: 

\begin{description}
\item[``BGS'' bar/bulge:] The bulge model is selected similar to 
[BGS]. It is a truncated power law bulge:
\bea
\rho\left( x,y,z\right) = b\cdot
\frac{e^{-a^2/a_m^2}}{a^3_m\eta\zeta\left(1+a/a_c\right)^q}
\eea
\item[``BGS'' disc:]
A double-exponential disc, without flaring [BGS]:
\bea
\rho\left( x,y,z\right) = d\cdot \left( e^{-|z|/z_0}/z_0 + \alpha 
e^{-|z|/z_1}/z_1\right) \cdot r_d e^{-r/r_d}
\eea
\item[``Kent'' disc:]
A double exponential disc, similar to the ``BGS'' disc. But now we 
include flaring, in spirit of the flaring
disc model of Kent et al. (1991). Inside of a galactocentric radius
$R_i=5\kpc$  the scale-height $h_z$ is constant. 
Outside of $R_i$ it rises linearly to the solar neighbourhood, where
the scale-height is $247\pc$. We also set the radial disc scale
length to the Kent et al. value of $r_d=3.001\kpc$ outside $R_i$. 
Inside $R_i$ the radial scale length is a fit variable. 

Thus we define for $r\leq 5\kpc$:
\begin{eqnarray*}
\rho\left( x,y,z\right) & = & d\cdot \left( e^{-|z|/z_0}/z_0 + \alpha 
e^{-|z|/z_1}/z_1\right) \\ 
& & \cdot r_d e^{-r/r_d} 
\end{eqnarray*}
for $r>5\kpc$, with $\sigma\equiv z_0 + \left(0.247\kpc - z_0 \right)\cdot
\frac{r/[{\tiny \kpc]}-5}{3}$:
\begin{eqnarray*}
\rho\left( x,y,z\right) & = & d\cdot \left( e^{-|z|/\sigma}/\sigma + \alpha 
e^{-|z|/z_1}/z_1\right) \\
&&  \cdot r_d e^{-r/3.001 {\tiny \kpc}} 
e^{5(3.001^{-1}-r_d^{-1}[{\tiny \kpc}])}
\end{eqnarray*}
We remark that this definition of the disc ensures $\rho\!\in\!C^0
\left(I\!\!R^3
\right)$, in particular that $\rho$ is continuous at $\{(x,y,z): r=
\sqrt[]{x^2+y^2}=5\kpc\}$. 
\end{description}
We remark that we assume a priori some of the parameters as fixed. 
These are the disc parameters $\alpha=0.27$ and scale-height $z_1=42\pc$, and the cusp  
parameters $a_c=0.1\kpc$ and $q=1.8$ [BGS].
We refer to Kent et al. (1991), [BM1] and [BGS] for a more detailed description of the  
models and their parameters. 

Using these model parts we define in table 1 
two models which we analyse. 
\begin{table*}
\begin{centering}
\begin{tabular}{|l|c|cc|}
\hline
Model & bulgemodel & discmodel &  \\
\hline
 & ``BGS-bulge''  & ``BGS-disc'' & ``Kent-disc'' \\
\hline
$\rho^{\textrm{noflare}}$ & X  & X & \\
$\rho^{\textrm{flare}}$ & X  & & X \\
\hline
\end{tabular}
\label{table1}
\caption{Combinations of the bulge/bar and disc model components to models.}
\end{centering}
\end{table*}

\subsection{The fitting algorithm}
The algorithm in order to find estimates for the parameters in the above
mentioned models is discussed in detail in [BM1] and only briefly 
summarised here.

Our task is to fit the de-reddened $COBE/DIRBE$ L-band surface brightness 
map $Y_{ij}=\omega^{\rm obs}(l_i,b_j)$ (Spergel et al. 1996), which is blurred by some 
random error $\varepsilon_{ij}$ at position $(l_i,b_j)$.
Particularly, an explorative data analysis shows that 
it is necessary to allow for a position dependent noise
${\rm Var}[\varepsilon_{i,j}]=\sigma^2_{i,j}$ (cf. [BM1]). 
The linear integral
operator ${\cal P}$ projects a three-dimensional luminosity distribution 
$\rho(x,y,z)$ to a surface-brightness distribution $\omega(l,b)$ at the sky,
viz:
\bea
\omega(l,b) = {\cal P}\left(\rho\right) (l,b)  =
\int_{0}^{\infty} \tilde\rho(r,l,b) dr,
\eea
with $\tilde\rho(r,l,b)\equiv\rho(x(r,l,b),y(r,l,b),z(r,l,b))$ where 
$\tilde\rho$ is defined in [BM1]. We assume
that  ${\cal P}$ is injective in a neighbourhood of $U= \{ \rho_{\vartheta} 
\left(\cdot\right)\}_{\vartheta  \in \Theta}$. 
This depends on a proper selection of the parametric
model $U$.  
Thus the problem to solve is to recover the MW luminosity density
$\rho^{MW}$ from the noisy 
integral equation $\omega^{\rm obs}(l_i,b_j)\equiv \omega^{MW}(l_i,b_j)+
\varepsilon_{ij}=
{\cal P}\left(\rho^{MW}\right) \left(l_i,b_j\right)+\varepsilon_{ij}$. Note
that $\omega^{MW}$ is the noise-free surface brightness distribution of the MW.

Following the method proposed in [BM1], let $\omega_{\vartheta}(l,b)\! =\!
{\cal P}\left(\rho_{\vartheta}\right)\left(l,b\right)$; 
$\vartheta\!\in\!\Theta$, and consider the transformed model
\bea
U_{\bf T}  = {{\bf T}U}=\left\{{\bf T}\omega_{\vartheta}(l,b)  
\right\}_{\vartheta \in\Theta},
\eea
with 
\bea
\left(\textrm{{\bf T}} \omega\right)(u,v) = \int\int \omega
(l,b) T((u,v), (l,b)) dldb.
\eea 
Here {\bf T} is a smoothing integral operator with kernel 
$T\left(\left(u,v\right),\left(l,b\right)\right)=\min\{u,l\}\cdot
\min\{v,b\}$; $(u,v),(l,b)\in I\!\!R^2$. Munk \& Ryumgaart (1999) have
shown that this smoothing kernel is a reasonable choice (cf. 
sect. 2 and [BM1]).   

According to sect. 2 we estimate the smoothed MW surface brightness $g^{MW}\left(u,v\right)=
\left({\bf T}\omega^{MW}\right)\left(u,v\right)$ from the noisy observations 
$Y_{ij}=\omega^{\rm obs}(l_i,b_j)$ as 
\bea
\hat{g}^{MW}(u,v)=\frac{1}{n\cdot m} \sum_{i=1}^n \sum_{j=1}^m \omega^{\rm obs}(l_i,b_j)
T((u,v),(l_i,b_j)),
\eea
and determine numerically the SMDE 
$\hat \vartheta_{\bf T}=\textrm{argmin}_{\vartheta\in
\Theta} ||\hat{g}^{MW}-{\bf T}\omega_{\vartheta}||^2_2$, where $||\cdot ||_2$ denotes
the usual $\textrm{L}^2$-norm. 
Finally the minimising value 
\bea
\hat{D}^2= ||\hat{g}^{MW}-{\bf T}\omega_{\hat{\vartheta}_{\bf T}}||^2
\eea
is computed. 

To this end we use the Marquardt-Levenberg-algorithm (Press et al., \cite{press})
for the minimisation in a two-step process:
\begin{description}
\item[1. Fitting of the disc parameters:]
In the first step we fit the disc parameters and the bulge
normalisation $b$, with the other bulge parameters fixed.
\item[2. Fitting of the bulge/bar parameters:]
In the second step we fix
the disc related parameters found in the first step (except for the
normalisation parameter $d$) 
and fit the bulge/bar parameters and $d$.
\end{description}

\begin{figure}
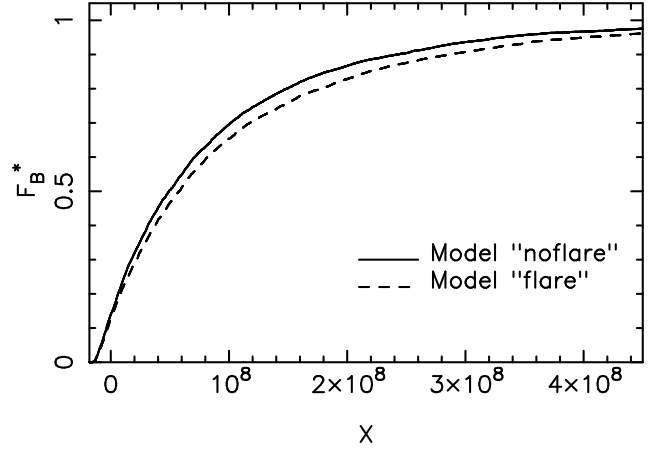

\getfig{plot1.eps}{6cm}
  \caption{Distribution $F_B^{\ast}$ of $X$ where 
$X=\sqrt[]{N}(\hat D^2_{\ast}-\hat D^2)$
for our models of the {\it COBE/DIRBE} 
L-band data. The solid line corresponds to model ``noflare'', the dashed line
to model ``flare''.}
\label{figfd}
\end{figure}

\begin{figure}
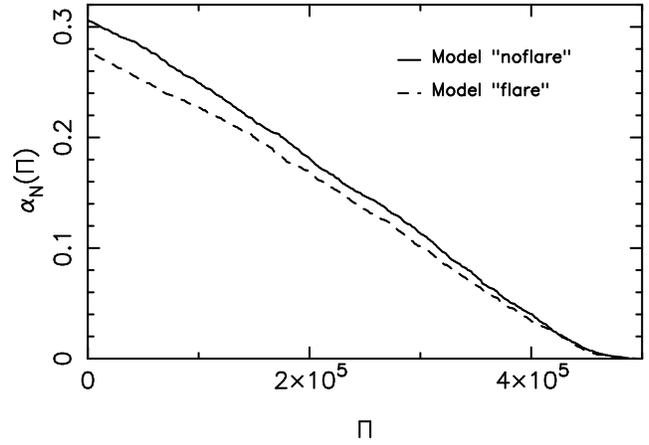

\getfig{plot5.eps}{6cm}
  \caption{The $p$-value curves $\alpha_N(\Pi)=F^{\ast}_B\left(
\sqrt{N}(\hat{D}^2-
\Pi)\right)$ for our two parametric models of the MW luminosity
density distribution. Model ``flare'' (dashed line) is better than model
``noflare'' (full line)
under the assumption that none of the models holds true for the data.}
\label{fighauptfig}
\end{figure}

\subsection{$P$-value curve analysis of our MW models}
Finally we analyse the MW models with our new method. 
To this end we first apply
the fitting algorithm (sect. 4.3) to both model ``flare'' and ``noflare''.
From this we obtain the best-fitting functions 
$f_{\hat\vartheta_{\bf T}}^{\rm flare/noflare}$, 
and the corresponding distances of the models from the data $\hat D^{2, 
{\rm flare/noflare}}$. Here superscript flare/noflare indicates that 
we determine this functions for model "flare" and (separately)  
for model "noflare".  
Then we perform the bootstrap analysis (sect. 2.2) 
for both model ``flare'' and ``noflare'', using the corresponding
best fit function and $\hat D^2$ for the respective model, and $B=5000$. 
This yields the empirical cumulative probability distribution functions \
$F_B^{\ast,{\rm flare/noflare}}$ of $X=\sqrt{N}\left(\hat D^{2, 
{\rm flare/noflare}}
-\Pi\right)$, and thus the functions $\alpha_N^{\rm flare/noflare}(\Pi)$. 
Fig. \ref{figfd} presents the resulting empirical distribution 
functions $F_{B}^{\ast, {\it flare/noflare}}$, and Fig. \ref{fighauptfig}
the $p$-value curves $\alpha_N^{\rm flare/noflare}(\Pi)$.

The last step in the analysis is to perform the graphical analysis 
of the $p$-value curves shown in Fig. \ref{fighauptfig}. For every
assumed distance $\Pi$ between model and the ``true'' function $f$ we find
more evidence for model ``flare'' than for model ``noflare'' (cf. sect. 3).
Therefore we are in the situation of graph $1$
in fig. \ref{figbeispielfig} and conclude that 
the $p$-value curve of model ``flare'' yields significantly more evidence for this
model than
that of model ``noflare''. Hence inclusion of flaring in the stellar disc 
improves the model. 
We can exclude that this conclusion
is due to over-fitting, because the entire $p$-value curve performs better.

However note that the present analysis provides much more information than
in [BM1], namely that there is {\it more statistical evidence for} ``flare'' as for
``noflare'' and {\it not only less evidence against} ``flare'' compared to ``noflare''.
This is because the method in [BM1] is based on the assumption that the model holds
(as essentially all classical goodness of fit procedures do),
and therefore ``only'' helps to decide whether the model should be rejected given the
observations. In contrast to this, for the new method proposed in this paper  
we assume that ``the model does not hold'', and estimate the probability that 
the distance between model and the ``true'' function $f$ 
is smaller than any assumed distance $\Pi$. Thus variation of $\Pi$
allows to find a (statistical) upper bound for the distance between model and data,
providing evidence {\it for} the model (within the limits of the
chosen distance between model and the true function $\rho$).

We  conclude that {\it inclusion of flaring in the double-exponential disc improves the
fit to the COBE/DIRBE L-band data.}  
This result is non-ambiguous, in particular
since the curve of model ``flare'' is below the curve of model ``noflare'' over
the entire scenario of possible distances $\Pi$ in fig. \ref{fighauptfig}. 

Determining $\Pi$ such that $\alpha_N(\Pi=D^2)\approx 0.1$ yields an estimate
for the distance between the best models with and without flaring disc component, 
respectively, and the true density distribution of the MW. 
We find $\Pi\approx 3\times 10^5$. This value can be considered as 
scientifically negligible because it is approximately equal to the
distance $\tilde D^2$ between the 
the best fitting parametric model of [BGS] and a variant thereof in which
the parameters have been changed in a random way by only $\approx 1\%$. 

\section{Final remarks and conclusions}
\subsection{Statistical Methodology}
We have suggested a method which allows to assess the validity of a regression
model at a controlled error rate. The error rate is fixed by deciding how large
$D^2$ may at the most be while still being considerable as scientifically negligible. 
Furthermore, several models can be compared.
This comparison is still possible if the models are parametrised by different
numbers of parameters since our method is sensitive to over-fitting of the data.
It is worthwhile to comment briefly on possible relationships to other
approaches. 
As pointed out by a referee, our approach is based on weighted least squares 
and hence, in a model with normal heteroscedastic errors, this is the 
maximum likelihood estimator (MLE). 
Note, however, that our approach does not require the assumption of a normal 
error, in general.

Other approaches in the literature 
are based on Bayesian ideas, e.g. model averaging where the
aim is to maximise the aposteriori probability
of a model $U_k$, say, given the observations $Y$, i.e. 
\be
P(T|Y)=\sum\limits_{k=1}^{l} P(T|U_k,Y)P(U_k|Y) 
\ee
where $l$ models are to be compared and $P(U_k|Y)$ denotes the posterior
probability of the model $U_k$ given $Y$. Here $T\hat{=}$ "pick the correct
model" (see Hoeting et al., 1999, DiCiccio et al., 1997). 
This approach is conceptually similar to ours, because it is based on
the idea that the decision in favor of or against a model should be investigated
under the full scenario of possible models. 
Bayesian model averaging aims for this by averaging, whereas we compare
all $P$-value curves among each other. However, in addition, we are in 
the position to decide whether the most appropriate model by such a rule
should be chosen at all. 
Interestingly Hoeting (p. 399) points out that such an investigation for
Bayesian model averaging would be of great interest. 
Another difficulty in Bayesian model selection consists in the determination
of priors. Observe, that our approach is based on a limit theorem,
which holds for any error distribution of $\varepsilon$, provided
${\rm Var}[\varepsilon]<\infty$. 
It would be 
important to investigate more closely these relationships, however this
is beyond the scope of this paper. 

\subsection{Flaring of the MW disc}
As an example application of our method we have compared a parametric model
of the MW luminosity distribution with a flaring vertical disc scale-height
$z_0$ with a model without flaring in the disc. We find that the model
with a flaring disc fits better the {\it COBE/DIRBE} L-band data than the model with 
constant $z_0$.  We conclude that the stellar disc flares outside some
inner radius $R_i$, which is significantly smaller than the
radius of the solar orbit $R_{\odot}$. 

Can young supergiant stars unrelated to the bulk of the stellar population, or 
polycyclic aromatic hydrocarbon $3.3\mu m$ or dust emission be responsible
for our result? Probably not, because Alard (2000) finds flaring of the
disc outwards of the solar orbit from {\it star count data}.
It seems improbable that near the solar circle the cause of the probably
same phenomenon changes. Also it is believed that the NIR luminosity 
probes the density of stars [BGS].

\section{Acknowledgments}
The authors are indebted to O. Gerhard and the referee E. Feigelson  
for many helpful comments. 
N. Bissantz acknowledges support by the Swiss Science Foundation
under grant number 20-56888.99, and thanks 
the University of Basel for their support, parts of this paper 
were written there. 
The source code can be obtained from the authors on request.
Please send an email to {\sl bissantz@math.uni-goettingen.de}

\end{document}